\documentclass[%
 aip,
 amsmath,amssymb,
 reprint,%
]{revtex4-1}

\usepackage{graphicx}
\usepackage{dcolumn}
\usepackage{bm}

\usepackage{CJK}
\usepackage[utf8]{inputenc}
\usepackage[T1]{fontenc}
\usepackage{mathptmx}
\usepackage{etoolbox}
\usepackage{xcolor}
\usepackage{soul}

\graphicspath{{./}{figures/}}

\makeatletter
\def\@email#1#2{%
 \endgroup
 \patchcmd{\titleblock@produce}
  {\frontmatter@RRAPformat}
  {\frontmatter@RRAPformat{\produce@RRAP{*#1\href{mailto:#2}{#2}}}\frontmatter@RRAPformat}
  {}{}
}%
\makeatother
\begin{document}
\begin{CJK*}{UTF8}{gbsn} 
\preprint{AIP/123-QED}

\title[polytropic solar wind model]{Acceleration of polytropic solar wind: Parker Solar Probe observation and one-dimensional model}
\author{Chen Shi (时辰)}
\email{cshi1993@ucla.edu}
\author{Marco Velli}%
 
\affiliation{ 
Department of Earth, Planetary, and Space Sciences, University of California, Los Angeles, CA 90095, USA
}%

\author{Stuart D. Bale}

\affiliation{
Physics Department, University of California, Berkeley, CA 94720-7300, USA}
\affiliation{Space Sciences Laboratory, University of California, Berkeley, CA 94720-7450, USA
}
\author{Victor R\'eville}
\affiliation{%
IRAP, Universit\'e Toulouse III - Paul Sabatier, CNRS, CNES, 31400 Toulouse, France
}

\author{Milan Maksimovi\'c}

\author{Jean-Baptiste Dakeyo}

\affiliation{%
LESIA, Observatoire de Paris, Universit e PSL, CNRS, Sorbonne Universit e, Universit e de Paris, 5 place Jules Janssen, 92195 Meudon, France
}%


\date{\today}

\begin{abstract}
The acceleration of the solar coronal plasma to supersonic speeds is one of the most fundamental yet unresolved problem in heliophysics. Despite the success of Parker's pioneering theory on an isothermal solar corona, the realistic solar wind is observed to be non-isothermal, and the decay of its temperature with radial distance usually can be fitted to a polytropic model. In this work, we use Parker Solar Probe data from the first nine encounters to estimate the polytropic index of solar wind protons. The estimated polytropic index varies roughly between 1.25 and 1.5 and depends strongly on solar wind speed, faster solar wind on average displaying a smaller polytropic index. We comprehensively analyze the 1D spherically symmetric solar wind model with polytropic index $\gamma \in [1,5/3]$. We derive a closed algebraic equation set for transonic stellar flows, i.e. flows that pass the sound point smoothly. We show that an accelerating wind solution only exists in the parameter space bounded by $C_0/C_g < 1$ and $(C_0/C_g)^2 > 2(\gamma-1)$ where $C_0$ and $C_g$ are the surface sound speed and one half of the escape velocity of the star, and no stellar wind exists for $\gamma > 3/2$. With realistic solar coronal temperatures, the observed solar wind with $\gamma \gtrsim 1.25$ cannot be explained by the simple polytropic model. We show that mechanisms such as strong heating in the lower corona that leads to a thick isothermal layer around the Sun and large-amplitude Alfv\'en wave pressure are necessary to remove the constraint in $\gamma$ and accelerate the solar wind to high speeds.
\end{abstract}

\maketitle
\end{CJK*}


\section{Introduction}\label{sec:intro}
Since the start of the space era, numerous human-made satellites that entered the interplanetary space have verified the existence of continuous, supersonic plasma flow, also known as the solar wind. Decades of satellite observations of the solar wind have revealed that solar wind interacts with the Earth's magnetosphere and injects a large amount of energy into the magnetosphere, causing various space weather events such as the magnetic storms and substorms that have great impacts on human society. Thus, understanding the solar wind, including how it is generated, is one of the most important topics in the field of space physics. 

Recent observations by Parker Solar Probe (PSP) show evidence of an accelerating solar wind close to the Sun\cite{shi2021alfvenic,sioulas2022magnetic}. Hence, it is now a good time to revisit the theory of solar wind generation.
The first theory of the formation of solar wind was established by Ref. [\onlinecite{parker1958dynamics}], who showed that with an isothermal and hot solar corona, the plasma is able to escape the solar gravity and becomes supersonic flow whose speed is similar to the in-situ observations. Ref. [\onlinecite{parker1964dynamicalI,parker1964dynamicalII,parker1965dynamical}] extend the theory to allow either a pre-defined temperature profile or a temperature that relates to the density through a static barometric law. In these early solar wind models, the only energy source is the efficient thermal conduction from the base of the solar corona. However, as the solar wind plasma is nearly collisionless, thermal conduction is only effective for the electron fluid\cite{hollweg1974electron,hollweg1976collisionless} but not the ions. Thus, other mechanisms are necessary for the acceleration and heating of solar wind. 

In the solar corona, various processes, e.g. magnetic reconnection \cite[e.g.][]{cargill2004nanoflare}, may provide a significant amount of energy, but these processes are important only at very low altitudes above the solar surface. It is now widely accepted that Alfv\'en waves are a promising power source of the solar wind, as large amplitude Alfv\'en waves are observed to be quasi-omni-present in the solar wind \cite[e.g.][]{belcher1971large}. In this scenario, outward propagating Alfv\'en waves are injected at the base of solar corona and are partially reflected because of the gradient of Alfv\'en speed \cite{heinemann1980non}. The reflected waves interact nonlinearly with the outward propagating waves, causing energy cascade from large to small scales (the turbulence energy cascade) \cite{kraichnan1965inertial}. The cascaded energy eventually dissipates through wave-particle interactions such as ion cyclotron resonance \cite{kasper2013sensitive} and Landau damping \cite{kobayashi2017three}, hence heats the plasma. In addition, the waves may directly accelerate the solar wind through the wave pressure gradient \cite{lionello2014validating} as the wave amplitude is large, especially around the Alfv\'en critical point. Many numerical works have shown that this Alfv\'en-wave-driven solar wind model is able to produce the observed fast solar wind \cite{verdini2009turbulence,shoda2018self,chandran2021approximate}.

The goal of the current study is to conduct a comprehensive analysis of the 1D one-fluid solar wind model with the Alfv\'en wave dynamics and heating in the lower corona properly approximated. We do not adopt a self-consistent Alfv\'en-wave-driven solar wind model \cite{verdini2009turbulence,lionello2014validating,reville2020role,chandran2021approximate}. Instead, we consider a polytropic solar wind model where the plasma thermal pressure $P$ and density $\rho$ obey the polytropic relation $P \rho^{-\gamma} = Const$ with a polytropic index $ \gamma $. Here $\gamma$ should be no larger than $5/3$ (adiabatic case) and no smaller than $1$ (isothermal case). For $\gamma =1$, the plasma can always gain sufficient heating, e.g. from thermal conduction, to maintain a constant temperature during its expansion. For $\gamma=5/3$, there is no heating of the plasma so that the internal energy must be consumed to support the work done by the plasma during its expansion, resulting in a cooling solar wind. A polytropic solar wind with $1<\gamma<5/3$ is in an intermediate state with finite heating, thus we can use it to approximate the heating effect from the Alfv\'en waves. 

We note that, the model studied here is one-fluid such that protons and electrons have identical number density, velocity, and thermal pressure. However, in the solar wind, electrons have complicated characteristics and dynamics compared with the protons, such as non-Maxwellian velocity distribution functions\cite{pierrard1999electron} and collisionless heat conduction\cite{hollweg1976collisionless}. Consequently, electrons have different temperature profiles from the protons\cite{newbury1998electron}, and an ambipolar electric field exists in the solar wind\cite{bervcivc2021ambipolar}. Thus, two-fluid models\cite{hartle1968two,tu1997two} are better in completely describing the dynamics of solar wind. Other important processes omitted in one-fluid model include the temperature anisotropy\cite{leer1972two} and kinetic instabilities such as mirror and firehose modes\cite{chandran2011incorporating}. Nonetheless, in this study, we focus on the one-fluid model because it can be solved in a semi-analytic way and some fundamental properties of the polytropic solar wind solution can be easily understood.

Many studies have been conducted to estimate the polytropic index of the solar wind proton. Helios data gives an average value $1.46$ which is independent of the solar wind speed \cite{totten1995empirical}. Measurements made at 1 AU confirm the speed-independence \cite{livadiotis2018long} and show that $\gamma$ is modulated by the solar activity \cite{nicolaou2014long}. Recent works \cite{nicolaou2020polytropic,huang2020proton} using PSP measurements, estimate the polytropic index to be close to or slightly smaller than $5/3$. 
There are some numerical-analytic studies of the polytropic stellar or solar wind. Ref. [\onlinecite{bondi1952spherically}] analyze the accretion problem in spherically symmetric geometry and discuss the property of solutions for polytropic plasma under stellar gravity and boundary conditions far from the star. Ref. [\onlinecite{parker1960hydrodynamic,demoulin2009temperature}] calculate the solar wind solution, assuming a non-self-consistent radial profile of either density or temperature. Recent study shows that the model of Ref. [\onlinecite{parker1960hydrodynamic}] produces good estimate of solar wind speed compared with the PSP data\cite{dakeyostatistical, halekas2022radial}. Ref. [\onlinecite{theuns1992spherically}] derive the closed form of the polytropic stellar wind equation but they focus on the solution of the shocked wind. A recent study by Ref. [\onlinecite{karageorgopoulos2021polytropic}] utilizes the complex plane strategy to solve the equation for polytropic stellar wind. 

In this study, we will use a different approach from that used by Ref. [\onlinecite{karageorgopoulos2021polytropic}] to solve the polytropic stellar wind model. The main point is to combine the integrated momentum equation (the Bernoulli's equation) with the polytropic relation and the mass-conservation equation to derive a single equation for the critical point. We will discuss how the polytropic index and the inner boundary temperature modify the wind solution. In addition, we approximate the effect of the coronal heating at low altitudes by assuming an isothermal layer at the bottom of the solar corona \cite{dakeyostatistical}. We also analyze the effect of a pre-defined force in the momentum equation. The paper is organized as follows: In Section \ref{sec:PSP_observation}, we show the statistical result, using PSP data from the first nine orbits, of the radial evolution of the proton temperature and estimate the polytropic index of the solar wind. In Section \ref{sec:polytropic_wind} we show the procedure to find the analytic-numerical solution of the polytropic stellar wind and do a comprehensive analysis of the characteristics of the model. In Section \ref{sec:conclusion}, we conclude the study.

\section{Parker Solar Probe observations}\label{sec:PSP_observation}
\begin{figure}
    \centering
    \includegraphics[width=\hsize]{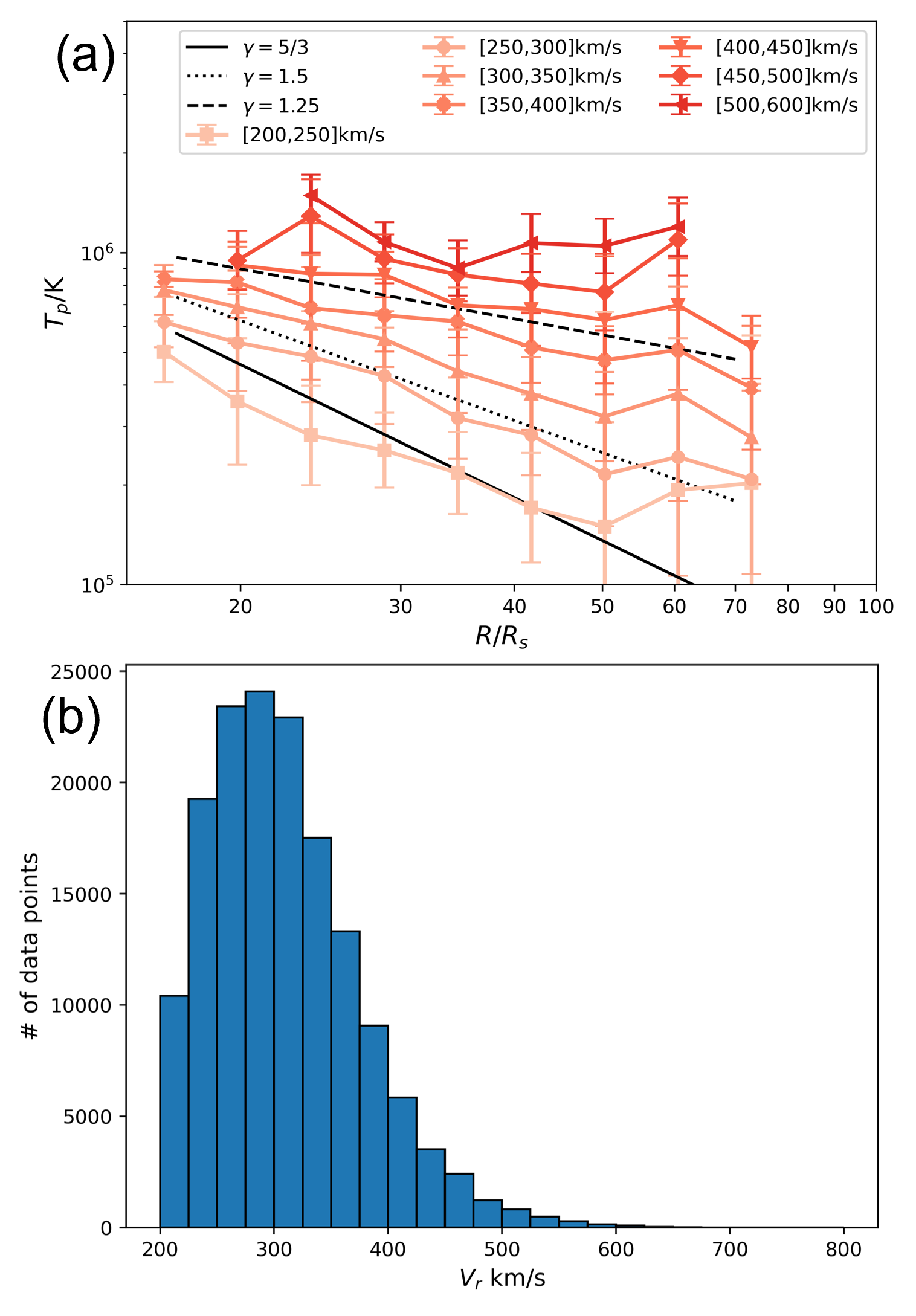}
    \caption{(a) Proton temperature ($T_p$) as a function of radial distance to the Sun. PSP/SPAN and PSP/SPC data from the first nine encounters are used. Curves with different colors correspond to different radial solar wind speeds as indicated in the legend. Squares and errorbars show the average values and standard deviations of the data binned in radial distances to the Sun. The black solid line shows $T_p \propto r^{-4/3}$, i.e. adiabatic cooling $\gamma=5/3$. The black dotted line shows $T_p \propto r^{-1}$, i.e. a polytropic solar wind with $\gamma=1.5$. The black dashed line shows $T_p \propto r^{-0.5}$, i.e. a polytropic solar wind with $\gamma=1.25$. (b): Number of data points as a function of solar wind speed range.}
    \label{fig:PSP_data}
\end{figure}

We use PSP proton data from the first nine encounters to estimate the polytropic index of the solar wind. Level-3 proton velocity and temperature data from SPAN-Ion (electrostatic analyzer) and SWEAP (Faraday cup) are used \cite{fox2016solar,kasper2016solar}. In FIG. \ref{fig:PSP_data} panel (a), we show proton temperature ($T_p$) as a function of radial distance to the Sun. 
Curves with different colors correspond to different radial solar wind speed ranges (written in the legend). Here we calculate the average values (squares) and standard deviations (errorbars) of the data binned in radial distance $r$. 
The black solid line shows $T_p \propto r^{-4/3}$, i.e. adiabatic cooling with $\gamma=5/3$. The black dotted line shows $T_p \propto r^{-1}$, i.e. a polytropic wind with $\gamma=1.5$. The black dashed line shows $T_p \propto r^{-0.5}$, i.e. a polytropic wind with $\gamma=1.25$. For all the speed ranges, $T_p$ roughly follows a power-law decay with $r$, implying a polytropic relation $T_p \propto r^{-2(\gamma - 1)}$ assuming the solar wind speed does not vary much with $r$. 
We note that, as PSP travels around the ecliptic plane, most of its observations are made inside slow solar wind. Panel (b) of FIG. \ref{fig:PSP_data} displays the probability distribution function of the solar wind speed from the first nine encounters. We can see that the peak of the distribution function is around $V_r=300$km/s and there are very few data points for $V_r>600$km/s. Thus, data points with $V_r>600$km/s are excluded in the statistics.
Panel (a) of FIG. \ref{fig:PSP_data} clearly shows that the polytropic index depends on the solar wind speed. Slower stream cools faster, with polytropic index around $1.5$, slightly smaller than the adiabatic index $5/3$, while faster stream cools slower, with polytropic index around $1.25$ or even smaller for speed larger than 450km/s. The observation indicates that the in-situ heating process, e.g. that from the turbulence cascade, is stronger in the faster solar wind stream.

\section{1D polytropic solar wind model}\label{sec:polytropic_wind}
In this section, we analyze the spherically symmetric 1D time-stationary solar wind model with purely radial velocity. In Section \ref{sec:isothermal}, we will briefly review the isothermal case \cite{parker1958dynamics}. In Section \ref{sec:polytropic_no_force}, we show in details how to determine the polytropic wind solution. Then we discuss the effect of an isothermal layer at the coronal base in Section \ref{sec:isothermal_layer} and finally we show the effect of the external force in Section \ref{sec:external_force}.

\subsection{Isothermal solar wind}\label{sec:isothermal}
\begin{figure}
    \centering
    \includegraphics[width=\hsize]{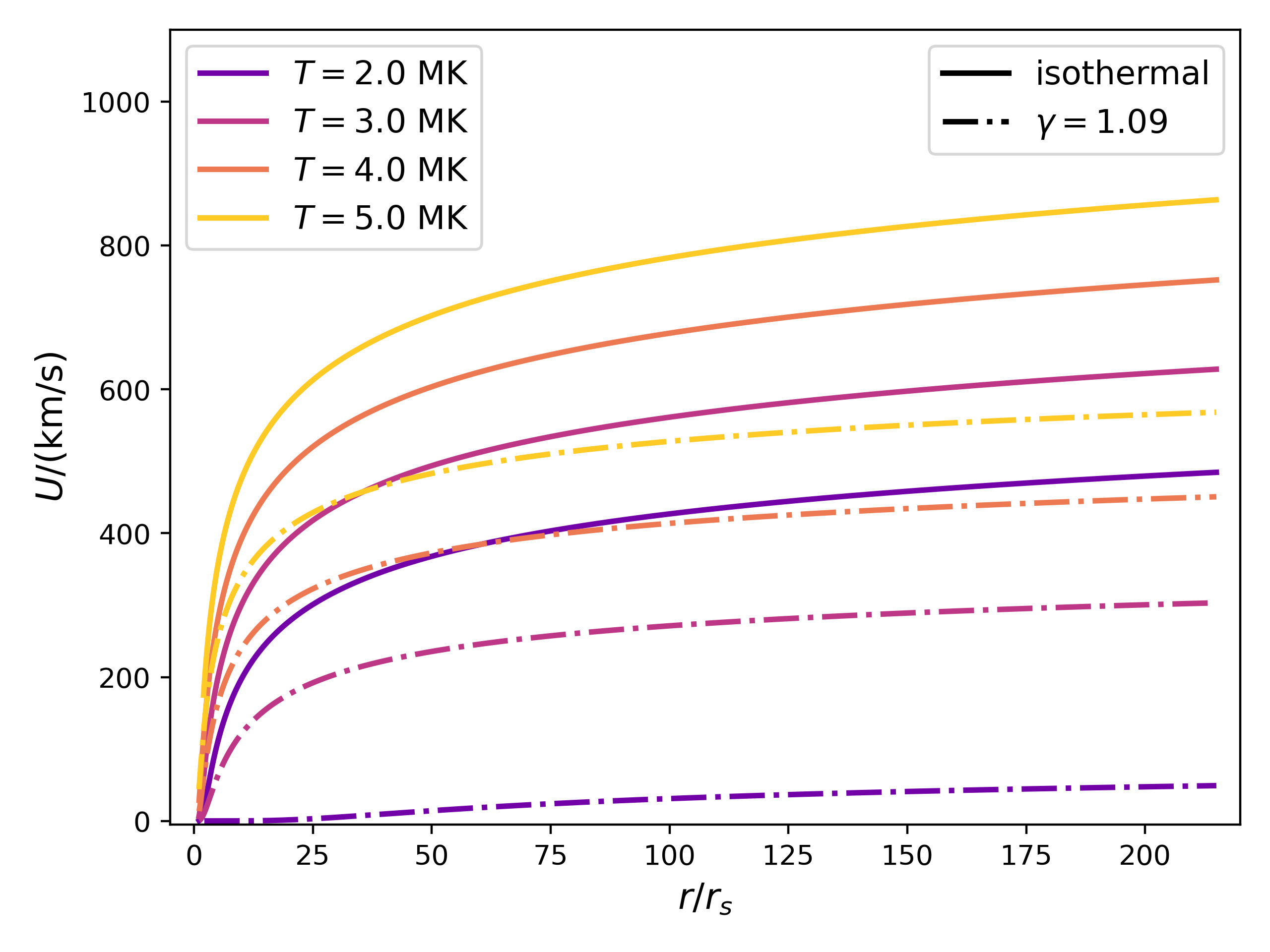}
    \caption{Solution of the isothermal (solid curves) and polytropic ($\gamma=1.09$, dashed-dotted curves) solar wind models for different temperatures at the solar surface ($r=r_s$).}
    \label{fig:U_R_isothermal_and_gamma109}
\end{figure}

As an introduction, we briefly review the 1D isothermal solar wind model with adiabatic index $\gamma = 1$, which was first analyzed by Ref. [\onlinecite{parker1958dynamics}]. Considering a flux tube with cross section area $A(r)$, the mass flux conservation gives 
\begin{equation}
\rho(r) V(r) A(r) = Const,
\end{equation}
where $\rho, V$ are the density and radial solar wind speed. The momentum equation is
\begin{equation}
    V \frac{dV}{dr} = - C^2 \, \frac{1}{\rho} \frac{d\rho}{dr} - \frac{GM}{r^2}
\end{equation}
with $C = \sqrt{k_B T/m_p}$ being the sound speed where $k_B$ is the Boltzmann constant, $m_p$ is the proton mass and $T$ is the sum of proton and electron temperatures (we have assumed the two temperatures are identical). In a spherically expanding solar wind ($A(r)=r^2$), the momentum equation can be re-arranged in the following form
\begin{equation}\label{eq:momentum_isothermal}
    \left( V - \frac{C^2 }{V} \right) \frac{dV}{dr} = C^2 \frac{2}{r} - \frac{GM}{r^2} 
\end{equation}
with the mass-conservation relation. The l.h.s. term indicates that if we require a solution of $V(r)$ that starts from a very small value at the inner boundary and increases to larger than the sound speed, the equation has a singular point
\begin{equation}\label{eq:rc_isothermal}
    r_c  = \frac{GM}{2C^2}
\end{equation}
at which $V(r)=C$. We can integrate equation (\ref{eq:momentum_isothermal}) from this critical (sound) point to any radial location $r$ and get the Bernoulli's equation
\begin{equation}\label{eq:isothermal}
\begin{aligned}
    \left( \frac{V^2}{C^2} - 1 \right) -  \ln \left(\frac{V^2}{C^2}\right) = & 4 \ln \left(\frac{r}{r_c}\right) + 4 \left( \frac{r_c}{r} - 1 \right) 
\end{aligned}
\end{equation}
In FIG. \ref{fig:U_R_isothermal_and_gamma109}, we show the solution of the isothermal wind model (equation (\ref{eq:isothermal})) for different coronal temperatures by solid curves. We note that there are actually two branches of solutions that pass through the critical point, but the other solution has $dV/dr < 0$, i.e. wind speed decreasing with distance and passing through the sound speed from above, thus it is not the solar wind solution we are interested in. However, it is worth noting that since the momentum equation is invariant under the transformation $V \rightarrow -V $, the decelerating solution represents an accretion flow\cite{bondi1952spherically,velli1994supersonic} which is likely related to the star formation\cite{mckee2007theory}. But as accretion is not important in the solar system, we do not analyze the decelerating solution in detail.

\subsection{Polytropic solar wind}\label{sec:polytropic_no_force}

In this section, we show how to self-consistently find the transonic solutions, i.e. solutions that pass through the sound point smoothly, with $\gamma > 1$. The polytropic case is more complicated than the isothermal case in that the critical point cannot be explicitly determined. A self-consistent analysis of the polytropic wind solution can be found in the textbook Ref. [\onlinecite{meyer2007basics}], while the approach presented in this section is different from that in Ref. [\onlinecite{meyer2007basics}].

The mass-conservation relation gives 
\begin{equation}\label{eq:rho}
    \rho(r) = \rho_0 \frac{ A_0 V_0}{A(r) V(r)}
\end{equation}
where $\rho_0$, $V_0$, and $A_0$ are the density, velocity, and cross section area at the inner boundary $r_0$. The polytropic relation gives $P\rho^{-\gamma} = Const$ or equivalently $T\rho^{-(\gamma-1)} = Const$. The momentum equation is 
\begin{equation}\label{eq:momentum}
    V \frac{dV}{dr} = - \frac{1}{\rho} \frac{dP}{dr} - \frac{GM}{r^2} + f(r)
\end{equation}
where $f(r)$ is the external force per unit mass. In this section, we set $f(r) = 0$. Using the polytropic relations, the pressure gradient term can be written as
\begin{equation}
    -\frac{1}{\rho } \frac{dP}{dr}  = -\frac{k_B}{m_p} \frac{\gamma}{\gamma -1} \frac{dT}{dr}
\end{equation}
Plug it into equation (\ref{eq:momentum}) and integrate from $r_0$ to $r$, we get
\begin{equation}\label{eq:mom_integral}
    \frac{1}{2} (V^2 - V_0^2) = - \frac{\gamma}{\gamma - 1} \frac{k_B}{m_p} (T-T_0) + GM\left(\frac{1}{r} - \frac{1}{r_0} \right) 
\end{equation}
The subscript ``0'' indicates quantities at the inner boundary $r_0$. The above equation includes two integral constants: $V_0$ and $T_0$. The temperature $T_0$ should be a given inner boundary condition but $V_0$ is not a free parameter and should be determined by the constraint that the solution $V(r)$ passes through the critical point smoothly. So, let's revisit the momentum equation (equation (\ref{eq:momentum})). We can write the pressure gradient term as
\begin{equation}
    - \frac{1}{\rho} \frac{dP}{dr} = C_s^2 \frac{1}{AV} \frac{d(AV)}{dr}
\end{equation}
where 
\begin{equation}\label{eq:Cs_and_C0_V0}
 C_s^2 = \frac{\gamma k_B T}{m_p} = C_0^2 \left(\frac{A_0 V_0}{AV} \right)^{\gamma-1} 
\end{equation}
is the square of \textit{local} sound speed and $C_0^2 = \gamma k_B T_0 / m_p$ is the square of sound speed at $r_0$. Plug the above relation into equation (\ref{eq:momentum}), we get
\begin{equation}\label{eq:eq_V}
    \left( V -  \frac{C_s^2}{V}\right) \frac{dV}{dr} = C_s^2 \frac{1}{A} \frac{dA}{dr} - \frac{GM}{r^2}
\end{equation}
One can easily show that for $\gamma=1$ and $A=r^2$, the above equation reduces to equation (\ref{eq:momentum_isothermal}). Equation (\ref{eq:eq_V}) implies that the location of the critical point $r_c$ and the velocity at the critical point $V_c$ satisfy
\begin{subequations}\label{eq:critical_point_eq_set}
\begin{equation}
    V_c - \frac{C_s^2}{V_c} = 0
\end{equation}
\begin{equation}
    C_s^2 \left(\frac{1}{A} \frac{dA}{dr}\right)_{r_c} - \frac{GM}{r_c^2} = 0
\end{equation}
\end{subequations}
or
\begin{equation}\label{eq:relation_Vc_and_rc}
    V_c^2 = C_s^2 = \frac{GM}{r_c^2} \times \left(\frac{1}{A} \frac{dA}{dr}\right)_{r_c}^{-1} 
\end{equation}
One more equation is needed to relate $V_0$ that appears in $C_s$ (equation (\ref{eq:Cs_and_C0_V0})) with $r_c$ and $V_c$. Writing the Bernoulli's equation (equation (\ref{eq:mom_integral})) at $r=r_c$ and using the relation $T_c = T_0 \left( A_0 V_0 / A_c V_c\right)^{\gamma-1}$
to substitute $T_c$, we get
\begin{equation}\label{eq:V0}
\begin{aligned}
    \frac{1}{2} (V_c^2 - V_0^2) = & - \frac{C_0^2}{\gamma - 1} \left[ \left( \frac{V_0A_0}{V_c A_c} \right)^{\gamma-1}   - 1\right] \\
    &+ GM \left(\frac{1}{r_c} - \frac{1}{r_0}\right)
\end{aligned}
\end{equation}

\begin{figure}
    \centering
    \includegraphics[width=\hsize]{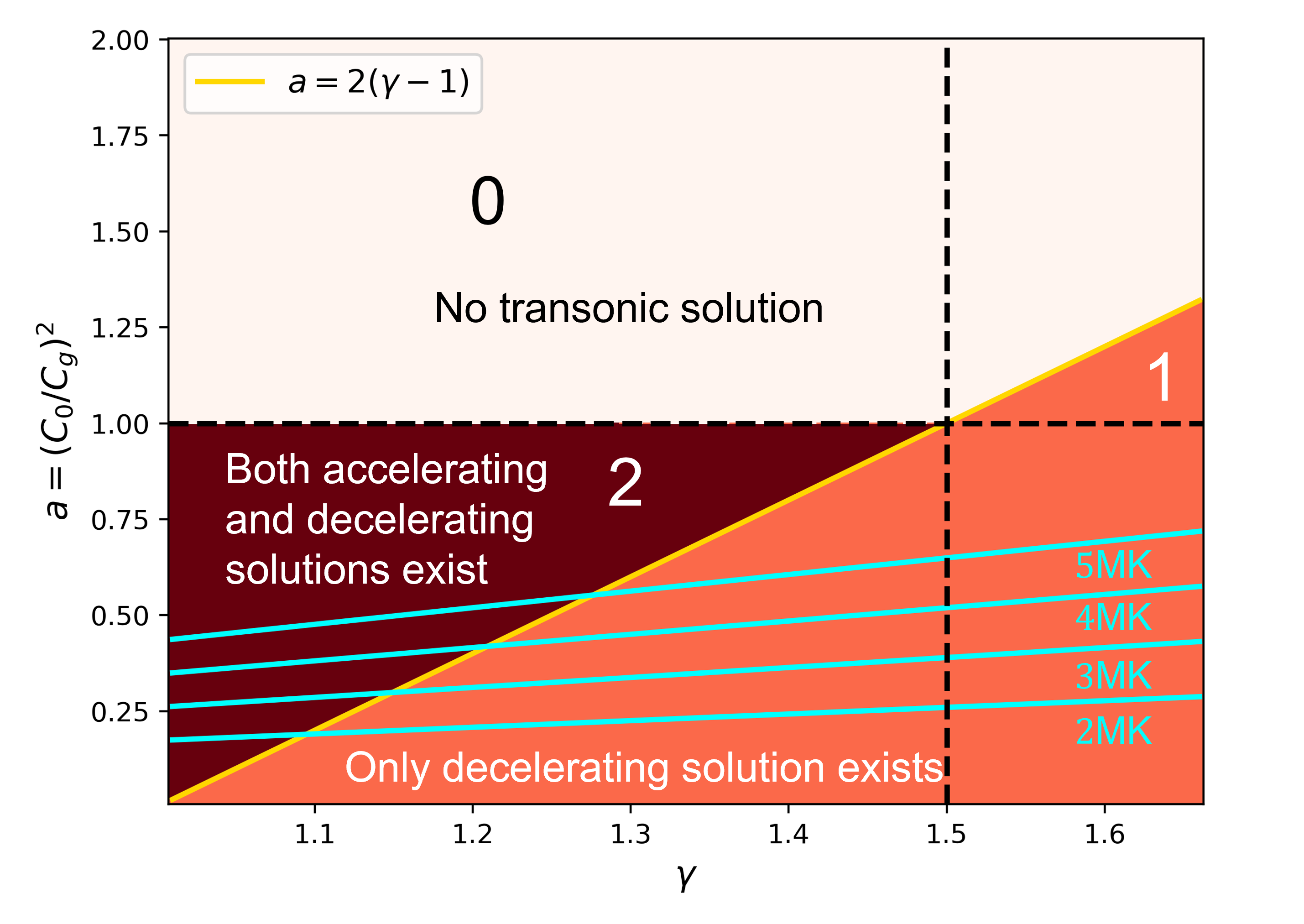}
    \caption{Phase diagram of the number of roots of equation (\ref{eq:sc_no_force}). Horizontal axis is polytropic index $\gamma$ and vertical axis is the parameter $a=(C_0/C_g)^2$ measuring the relative strength of the gravitational field. Dark red region has two roots, orange region has one root, and the white region has no root. The yellow line shows $a=2(\gamma - 1)$. The horizontal dashed line marks $a=1$ and the vertical dashed line marks $\gamma=3/2$. The cyan lines show the realistic parameters of the Sun with $r_0=r_s$ and varying inner boundary temperature $T_0$.}
    \label{fig:Nroots_sc_general}
\end{figure}

Equations (\ref{eq:critical_point_eq_set}a), (\ref{eq:relation_Vc_and_rc}), and (\ref{eq:V0}) form a closed equation set:
\begin{subequations}\label{eq:three_equation_set}
\begin{equation}\label{eq:three_equation_critical_speed_and_sound_speed}
    V_c^2 = C_s^2 = C_0^2 \left(\frac{A_0 V_0}{A_c V_c} \right)^{\gamma-1}
\end{equation}
\begin{equation}\label{eq:three_equation_critical_speed_and_gravity}
    V_c^2 = \frac{GM}{r_c^2} \times \left(\frac{1}{A} \frac{dA}{dr}\right)_{r_c}^{-1}
\end{equation}
\begin{equation}\label{eq:three_equation_integrated_mom_eq}
\begin{aligned}
     \frac{1}{2} (V_c^2 - V_0^2) = & - \frac{C_0^2}{\gamma - 1} \left[ \left( \frac{V_0A_0}{V_c A_c} \right)^{\gamma-1}  - 1\right] \\
     &+ GM \left(\frac{1}{r_c} - \frac{1}{r_0}\right)
\end{aligned}
\end{equation}
\end{subequations}
from which we can solve $V_0$, $V_c$, and $r_c$ simultaneously. Here we note that this three-equation model can be easily extended to a multi-temperature fluid, e.g. a wind where the ion and electron have different temperatures and polytropic indices\cite{dakeyostatistical}. The only modifications are the definition of the sound speed and the pressure gradient term in equation (\ref{eq:three_equation_integrated_mom_eq}) as we need to sum up the contributions from all species. For simplicity, we assume a single-fluid solar wind throughout this study.

We consider a radially-expanding solar wind ($A(r) = r^2$).
Then from equation (\ref{eq:three_equation_critical_speed_and_gravity}), we get
\begin{equation}\label{eq:relation_Vc_and_Cg}
    V_c^2 = \frac{GM}{2r_c} =  \frac{C_g^2}{s_c}
\end{equation}
where we have defined the normalized radius $s = r/r_0$ and
\begin{equation}\label{eq:cosmic_speed}
    C_g = \sqrt{\frac{GM}{2r_0}},
\end{equation} 
which is one half of the escape velocity. We note that $r_0$ does not necessarily equal to the solar radius $r_s$. As will be discussed in Section \ref{sec:isothermal_layer}, $r_0$ could be the outer radius of the isothermal layer formed due to large heating or thermal conduction in the lower corona \cite{dakeyostatistical}.
Plug equation (\ref{eq:cosmic_speed}) in equation (\ref{eq:three_equation_critical_speed_and_sound_speed}), we get
\begin{equation}\label{eq:relation_V0_and_rc}
    V_0^2 = C_g^2  \left( \frac{C_g}{C_0} \right)^{\frac{4}{\gamma-1}}  s_c^{\frac{3\gamma-5}{\gamma-1}} 
\end{equation}
With equations (\ref{eq:relation_Vc_and_Cg}) and  (\ref{eq:relation_V0_and_rc}), we can eliminate $V_0$ and $V_c$ in equation (\ref{eq:three_equation_integrated_mom_eq}) and get a single equation for $s_c$:
\begin{equation}\label{eq:sc_no_force}
\begin{aligned}
    \frac{1}{2} \left[ \left( \frac{C_g}{C_0}\right)^{\frac{4}{\gamma-1}} s_c^{\frac{2(2\gamma-3)}{\gamma-1}} - 1 \right] + & \frac{1}{\gamma - 1} \left( \frac{C_0^2}{C_g^2} s_c - 1 \right) \\
    &= 2(s_c-1)
\end{aligned}
\end{equation}
This is an algebraic equation, so we can numerically find all the roots $s_c(\gamma,C_0/C_g)$. 
An alternative form of the equation can be found in Ref. [\onlinecite{reville2016vents}].
In Appendix \ref{sec:append_Nroot}, we describe in detail how to determine the number of roots of equation (\ref{eq:sc_no_force}) with the knowledge of $C_0/C_g$ and $\gamma$. In FIG. \ref{fig:Nroots_sc_general} we show the phase diagram of the number of roots of equation (\ref{eq:sc_no_force}) on the $(C_0/C_g)^2$-$\gamma$ plane. Dark red region has two roots, orange region has one root, and the white region has no root. The yellow line is $(C_0/C_g)^2 = 2(\gamma -1)$. 
We have numerically verified, by comparing $V_c$ and $V_0$, that if there is one root, only the decelerating solution exists, i.e. solar wind solution can only be found in the dark red region, bounded by the two lines $C_0/C_g=1$ and $(C_0/C_g)^2 > 2(\gamma-1)$. 
Naturally, $\gamma < 3/2$ must also be satisfied as the two lines cross at $\gamma = 3/2$. Early study \cite{parker1960hydrodynamic} using asymptotic analysis points out that physically meaningful solar wind solution exists only with $\gamma < 3/2$ and $1 \gg (C_0/C_g)^2 > (\gamma-1)/2\gamma $. Our result gives a more precise constraint on $(C_0/C_g)^2$. The cyan lines in FIG. \ref{fig:Nroots_sc_general} mark the realistic parameters of the Sun with $r_0 = r_s$ and varying temperature at the inner boundary $T_0$. One can read that for $T_0=2$MK solar wind does not exist for $\gamma \gtrsim 1.1$, and even for a very hot corona with $T_0=5$MK, the solar wind does not exist for $\gamma \gtrsim 1.3$.

\begin{figure}
    \centering
\includegraphics[width=\hsize]{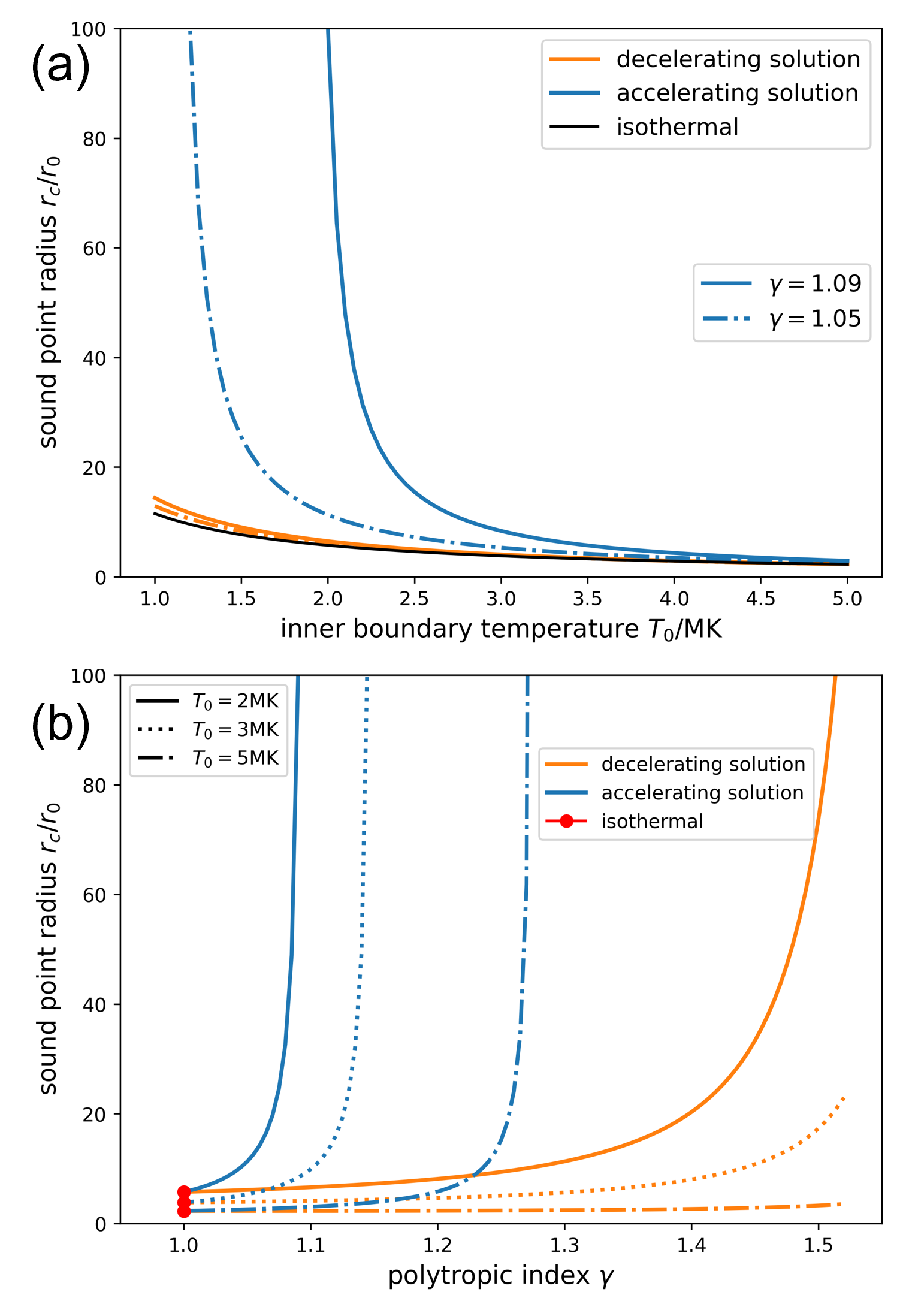}
    \caption{(a) Location of the sound point $r_c/r_0$ ($r_0=r_s$) as a function of the inner boundary temperature $T_0$. Blue curves are the accelerating (wind) solution and orange curves are the decelerating solution. Solid curves are $\gamma=1.09$ and dashed-dotted curves are $\gamma=1.05$. Black curve is the isothermal case. (b) Location of the sound point as a function of the polytropic index $\gamma$ with fixed inner boundary temperature. Solid curves are $T_0=2$MK, dotted curves are $T_0=3$MK, and dashed-dotted curves are $T_0=5$MK. Red dots mark the isothermal cases.}
    \label{fig:critical_radius}
\end{figure}

Figure \ref{fig:critical_radius} shows $s_c$ as a function of $T_0$ with fixed $\gamma$ (panel (a)) and $s_c$ as a function of $\gamma$ with fixed $T_0$ (panel (b)). In this plot, we have set $r_0=r_s$. Blue curves show the accelerating (solar wind) solution and orange curves show the decelerating solution. The black curve in panel (a) and the red dots in panel (b) correspond to an isothermal plasma ($\gamma = 1$). Different from the isothermal case where the accelerating and decelerating solutions pass through the same critical point, the two branches of solutions with $\gamma > 1$ have different critical radii because the temperature depends on the wind speed profile. As $T_0$ decreases and $\gamma$ increases, the critical radii for both the two branches of solutions increase, and $r_c$ of the accelerating solution increases much faster than that of the decelerating solution. Consistent with FIG. \ref{fig:Nroots_sc_general}, one can read from panel (b) of FIG. \ref{fig:critical_radius} that $r_c$ for the accelerating solution diverges to extremely large values (actually infinity) as $\gamma$ approaches certain critical values, corresponding to the cross points between the cyan and yellow lines in FIG. \ref{fig:Nroots_sc_general}.

\begin{figure}
    \centering
\includegraphics[width=\hsize]{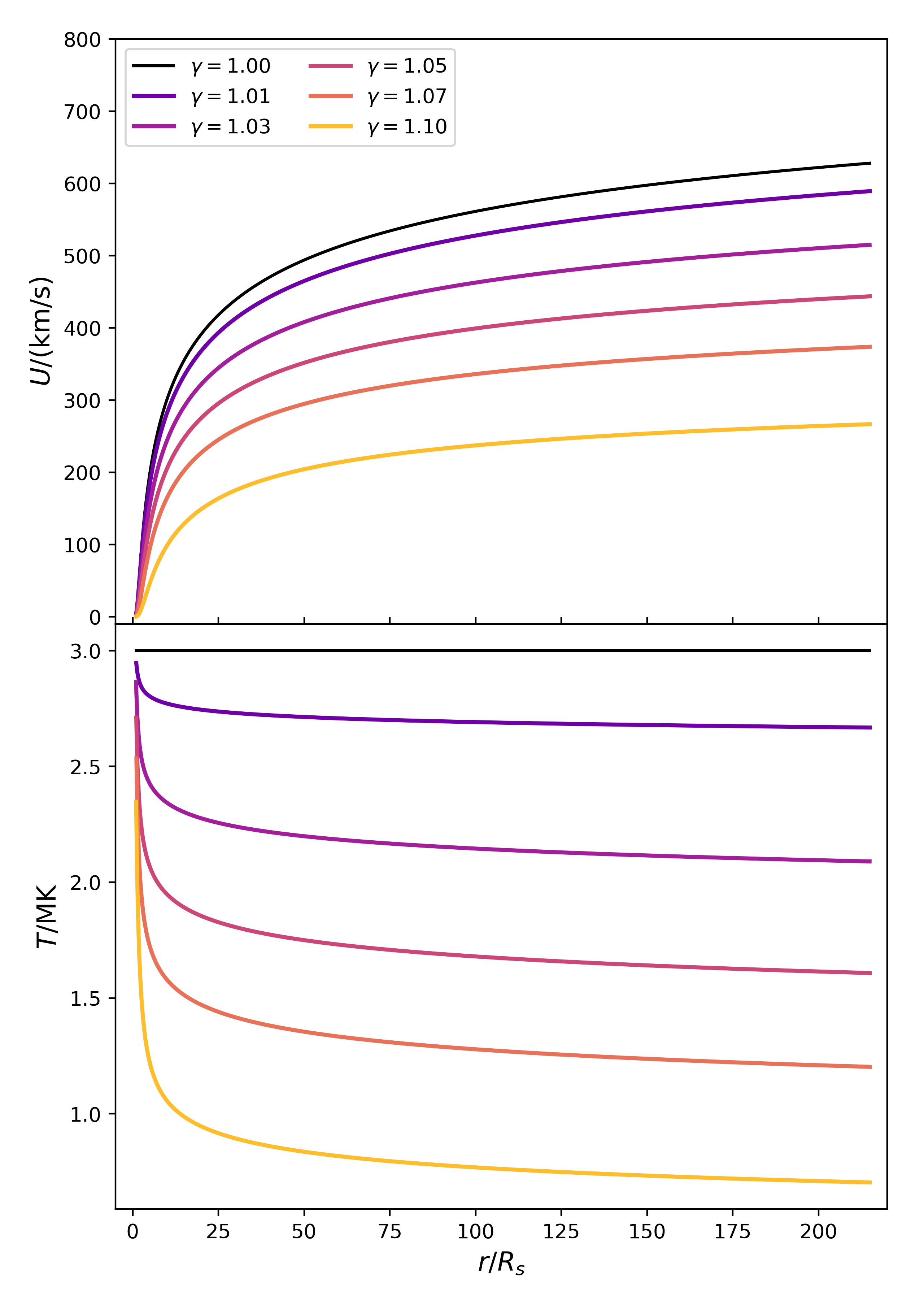}
    \caption{Solar wind solution for $T_0=3$MK and varying $\gamma$. Top panel shows the wind speed and bottom panel shows the temperature.}
    \label{fig:U_T_R_diff_gamma}
\end{figure}

After $s_c$ is solved, we can easily calculate $V_c$ from equation (\ref{eq:three_equation_critical_speed_and_gravity}) and then integrate the momentum equation from $r_c$:
\begin{equation}
\begin{aligned}
    \frac{1}{2} (V^2 - V_c^2) = & - \frac{V_c^2}{\gamma - 1} \left[ \left( \frac{V_c s_c^2}{V s^2} \right)^{\gamma-1} - 1 \right] \\ &
    + 2 C_g^2 \left( \frac{1}{s} - \frac{1}{s_c} \right)
\end{aligned}
\end{equation}
to acquire the profile of $V(r)$. We note that, the above equation gives two branches of $V(r)$ starting from one critical point, but only one branch satisfies the given inner boundary condition. This is similar to the early work on polytropic accretion flow\cite{bondi1952spherically}, which shows that there are two branches of solutions that cross the critical point but only one of them satisfies the boundary condition at infinity. In Figure \ref{fig:U_R_isothermal_and_gamma109}, we plot $V(r)$ for $\gamma=1.09$ and varying $T_0$ in dashed-dotted curves. Compared with the isothermal case, the solar wind speed drops significantly even though $\gamma$ is only slightly larger than 1. In Figure \ref{fig:U_T_R_diff_gamma}, we show the profiles of the solved solar wind speed (top) and the corresponding temperature (bottom) for a fixed inner boundary temperature $T_0=3$MK and different values of $\gamma$. We see that the wind speed decreases very rapidly with the increasing $\gamma$. 

In conclusion, results from this section show that a simple polytropic wind model cannot explain the observed solar wind speed with the polytropic index deduced from in-situ data as shown in Section \ref{sec:PSP_observation}. In the following two sections, we will discuss two mechanisms that possibly contribute to solve the problem.

\subsection{Isothermal layer}\label{sec:isothermal_layer}
In the lower corona, heating mechanisms such as magnetic reconnection and high thermal conduction may prevent the plasma temperature from decaying much. Hence, one way to overcome the difficulty that solar wind solution does not exist with large polytropic index is to assume there is an isothermal zone \cite{dakeyostatistical} with radius $r_{iso}>r_s$ such that $T(r\leq r_{iso}) \equiv T_0$. In the recent study by Ref. [\onlinecite{dakeyostatistical}], this ``iso-poly'' solar wind model is solved numerically with quite thick isothermal layers such that the critical point always falls inside the isothermal layer. In this section, we discuss the model in more details and analyze the case when the critical point is outside the isothermal layer. Especially, we point out that, due to the discontinuity in $\gamma$ at the interface between the isothermal and polytropic layers, the values taken by $r_{iso}$ present a radial interval (around the critical point), for which iso-poly solar wind solutions do not exist.  

\begin{figure}
    \centering
    \includegraphics[width=\hsize]{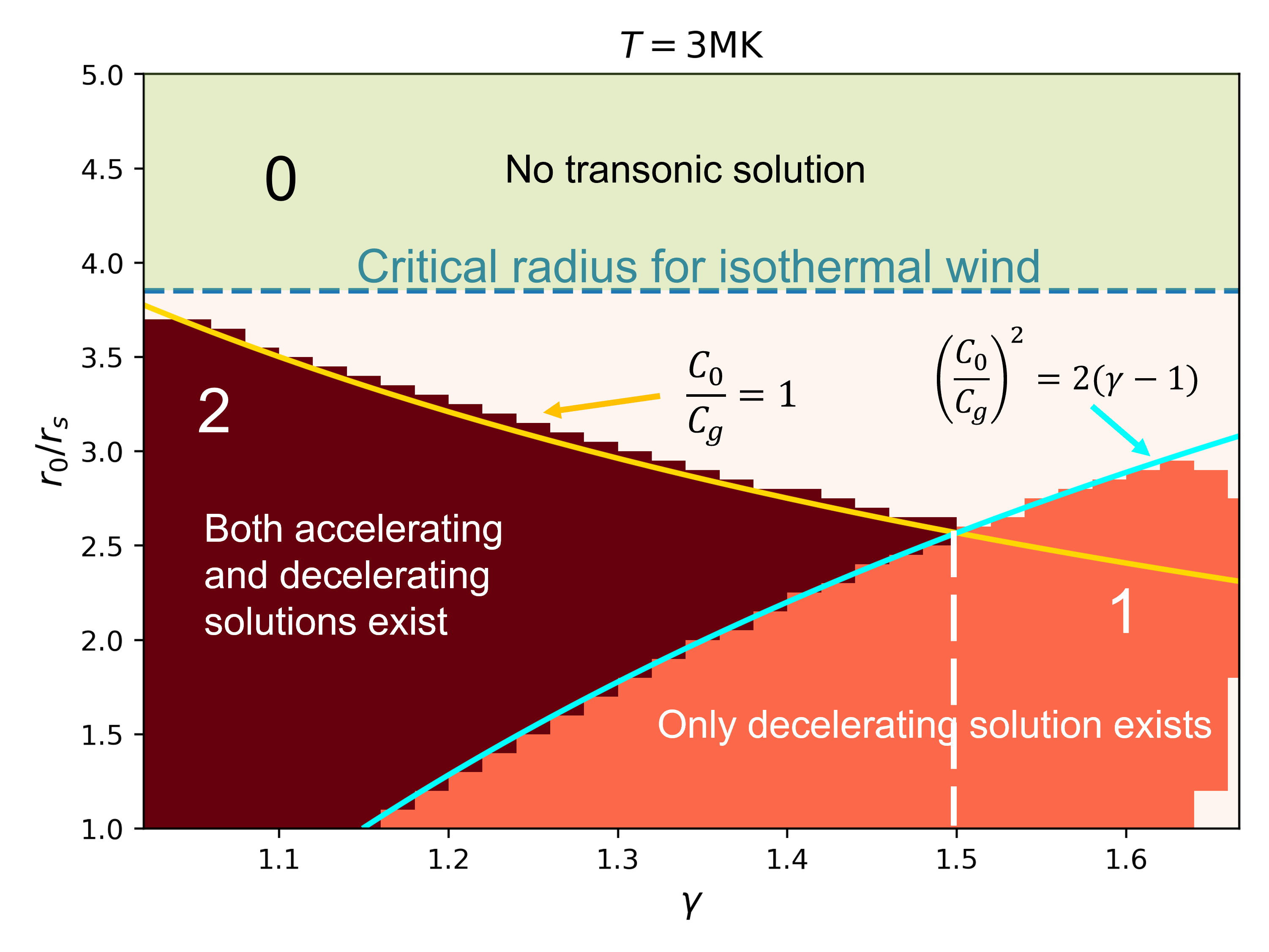}
    \caption{Phase diagram of the solutions to equation (\ref{eq:three_equation_set}) on the $r_0/r_s - \gamma$ plane, with a base temperature $T_0=3$MK. Here the numbers of solutions are determined by numerically searching roots of equation (\ref{eq:sc_no_force}) in the range $r\in[1,20000]r_0$. The dark red region has both accelerating and decelerating solutions, the orange region has only decelerating solution, and other regions have no solution. The yellow curve marks $C_0/C_g=1$ and the cyan curve marks $(C_0/C_g)^2 = 2(\gamma - 1)$. The horizontal dashed line marks the critical radius for the isothermal wind $r_{c,iso}$.}
    \label{fig:phase_diagram_isothermal_base}
\end{figure}

For the polytropic layer, if the critical point is still inside the layer ($r_{c}>r_{iso}$), the procedure described in the prior section to solve equation (\ref{eq:three_equation_set}) remains exactly the same. We only need to set the inner boundary at $r_0 = r_{iso}$ and re-define $C_g$ using equation (\ref{eq:cosmic_speed}). In FIG. \ref{fig:phase_diagram_isothermal_base}, we show the phase diagram of the solutions to equation (\ref{eq:three_equation_set}) on the $r_0/r_s - \gamma$ plane, with a base temperature $T_0=3$MK. The plot is similar to FIG. \ref{fig:Nroots_sc_general} but here the number of solutions is determined by numerically searching roots of equation (\ref{eq:sc_no_force}) in the range $r\in[1,20000]r_0$ instead of using the method described in Appendix \ref{sec:append_Nroot}. Thus it also serves as a verification of the method described in Appendix.
In this plot, the yellow curve marks $C_0/C_g=1$ and the cyan curve marks $(C_0/C_g)^2 = 2(\gamma - 1)$. 
The horizontal dashed line marks the critical radius for the isothermal wind $r_{c,iso}$. Note that the yellow curve intersects with the horizontal line at $\gamma = 1$ (equation (\ref{eq:rc_isothermal})). 

\begin{figure}
    \centering
    \includegraphics[width=\hsize]{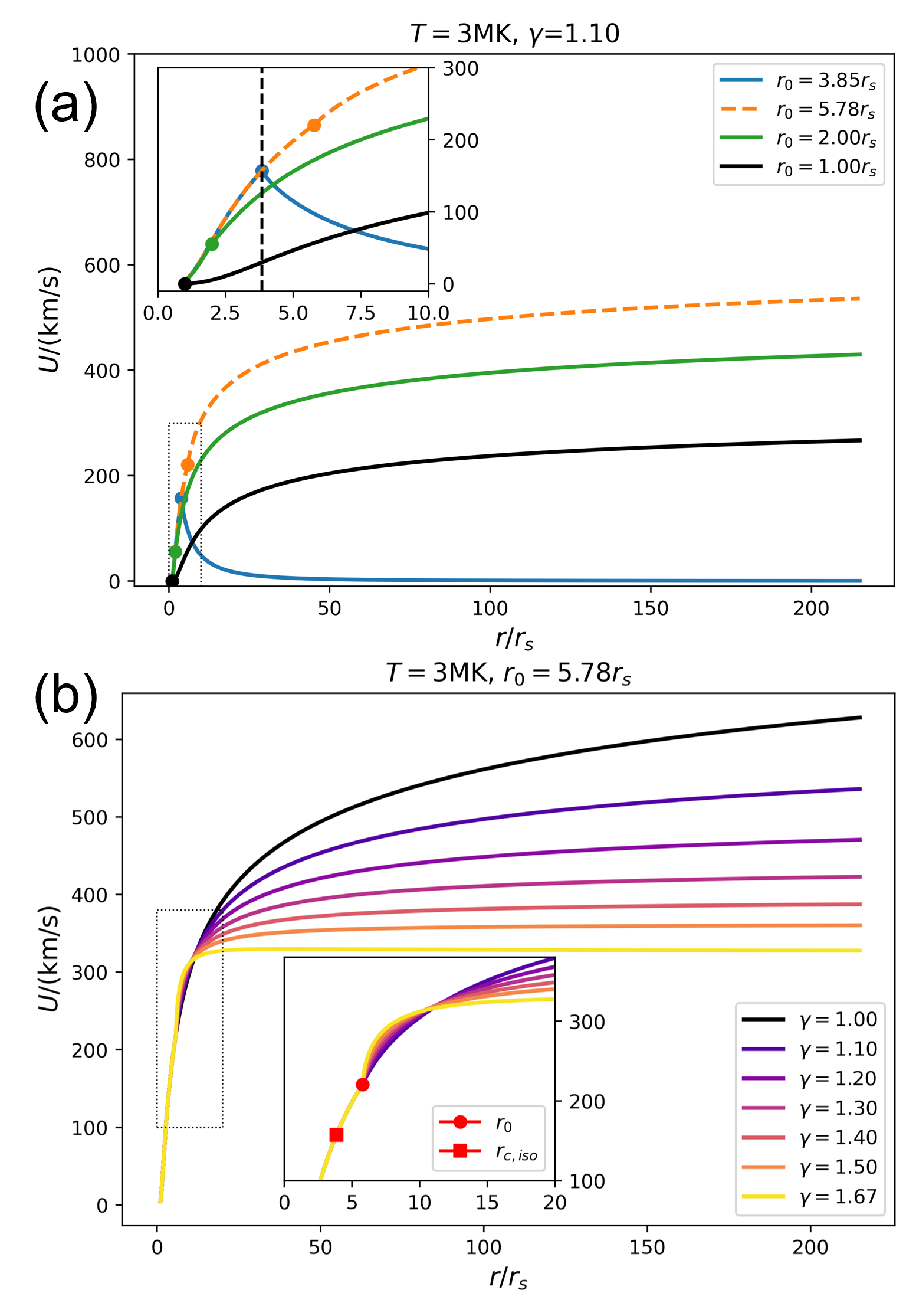}
    \caption{(a) Profiles of solar wind speed for $T_0=3$MK, $\gamma=1.1$ and varying radius of the isothermal layer $r_{0}$ ($r_{iso}$). Black curve is $r_0=r_s$, i.e. no isothermal layer. Green curve is $r_0=2r_s$. Blue curve is $r_0=3.85r_s=r_{c,iso}$ where $r_{c,iso}$ is the critical radius in isothermal case. Orange dashed curve is $r_0=5.78r_s=1.5r_{c,iso}$. The circles mark the locations of $r_0$. Embedded plot is a blow-up of the dotted rectangle. The dashed vertical line marks $r_{c,iso}=3.85r_s$. (b) Profiles of solar wind speed for $T_0=3$MK, $r_0=5.78r_s$ and varying $\gamma$. Dark to light colors correspond to $\gamma$ from $1$ to $5/3$. Embedded plot is a blow-up of the dotted rectangle. Red circle marks $r_0$ and red square marks $r_{c,iso}$.}
    \label{fig:U_R_base}
\end{figure}

If $(r_{iso},\gamma)$ falls in the red region, i.e. the iso-poly critical point is in the polytropic layer. We can first calculate the solar wind solution in the polytropic layer following the procedure in Section \ref{sec:polytropic_no_force}. We will also acquire the solar wind speed at the inner boundary $r_{iso}$. As the wind speed must be continuous, $V(r_{iso})$ serves as the upper boundary condition for the isothermal layer. Thus, we can then integrate the isothermal momentum equation back from $r_{iso}$ to $r_s$. In panel (a) of FIG. \ref{fig:U_R_base}, we show profiles of solar wind speed for $T_0=3$MK, $\gamma=1.1$ and varying radius of the isothermal layer $r_{0}$ ($r_{iso}$). Black curve is $r_0=r_s$, i.e. no isothermal layer, and green curve is $r_0=2r_s$, a case that the critical point falls in the polytropic layer. The circles mark the locations of $r_0$. One can see that raising the height of the isothermal layer indeed increases the solar wind speed. Besides, FIG. \ref{fig:phase_diagram_isothermal_base} shows that changing $r_{iso}$ also changes the range of $\gamma$ in which a solar wind solution can be found, though the maximum $\gamma$ value allowed is always $3/2$ with $r_{c}$ in the polytropic layer.

If we set $r_{iso} > r_{c,iso} $ (in the green region on FIG. \ref{fig:phase_diagram_isothermal_base}), no transonic solution exists in the polytropic layer and the critical point, if exists, falls inside the isothermal layer such that $r_c = r_{c,iso}$. In this case, we can first determine the solution in the isothermal layer, which is simply the classic Parker's model (Section \ref{sec:isothermal}). The solution then gives $V(r_{iso})$, which serves as the inner boundary condition for the polytropic layer. In panel (a) of FIG. \ref{fig:U_R_base}, the orange dashed curve is $r_{0}=5.78r_s=1.5r_{c,iso}$. We get a higher wind speed than the cases $r_0=2r_s$ and $r_0=r_s$. 
However, one limitation appears: the difference between the isothermal sound speed and the polytropic one, does not ensure that the wind speed continues to increase beyond $r_0$. For instance, let us consider the following scenario: $r_0$ is only slightly larger than (or equal to) $r_{c,iso}$ so that the isothermal layer gives $V(r_0) \gtrsim \sqrt{k_B T /m_p} = C$. However, for the polytropic solar wind, we have $ C_0 = \sqrt{\gamma k_B T / m_p} > \sqrt{k_B T /m_p} $, which may lead to $V(r_0) < C_0$ in the polytropic layer, i.e. the supersonic solar wind suddenly becomes subsonic across $r=r_0$ for this case.
In other words, since the wind speed $V(r)$ does not overcome $C_0$ before going out of the isothermal layer, there is no possible supersonic expansion in the polytropic layer (no overlap between the green and red regions in FIG. \ref{fig:phase_diagram_isothermal_base}), and the flow can not remain supersonic above $r_0$. Then one will get either a blow-up solution with $dV/dr \rightarrow \infty$ at some point or a ``breeze'' with $V(r \rightarrow +\infty) \rightarrow 0$. The blue curve on the panel (a) of FIG. \ref{fig:U_R_base} illustrates the case of a "breeze" solution with $r_0=3.85r_s=r_{c,iso}$, for which beyond $r_0$, the accelerating solar wind suddenly starts decelerating and the wind speed tends to zero at large $r$.
Another limitation of the iso-poly model appears when we consider the white parameter space (region between the yellow ($C_0/C_g=1$) curve and the $r_{c,iso}$ line) of FIG. \ref{fig:phase_diagram_isothermal_base}. 
Within this parameter space, the isothermal layer is thinner than $r_{c,iso}$ so that the solar wind cannot be accelerated to a supersonic speed below $r_{iso}$. Meanwhile, the polytropic layer does not have an accelerating transonic solution. Thus, for this parameter region, we can not find a transonic iso-poly solution.


In panel (b) of FIG. \ref{fig:U_R_base}, we plot iso-poly wind speed profiles for $T_0=3$MK, $r_0=5.78r_s=1.5r_{c,iso}$ and different values of $\gamma$. The black curve represents the isothermal case, and dark to light colors correspond to increasing values of $\gamma$. The red circle marks $r_{iso}$ and the red square marks $r_c$(=$r_{c,iso}$). Below $r_{iso}$, the solutions are exactly the same for all the $\gamma$. Indeed, within the isothermal layer, all the iso-poly models have the same initial temperature, and they passe through the same sonic point. Slightly above $r_{iso}$, a larger $\gamma$ value leads to a larger acceleration of the wind on a short radial distance, but it also gives a smaller asymptotic wind speed. We note that, in this case (solutions in the green region of FIG. \ref{fig:phase_diagram_isothermal_base}), $\gamma$ can be larger than $3/2$.

In conclusion, the iso-poly model is able to produce transonic wind solutions under either of the following two conditions: 
$(i)$  $r_{iso} > r_{c,iso}$ and $V(r_{iso}) > (C, C_0)$, i.e. the wind is supersonic and faster than both sound speeds as it enters the polytropic layer. This case corresponds to the orange curve in panel (a) of FIG. \ref{fig:U_R_base}.
$(ii)$ ($\gamma$, $r_{iso}$) falls in the dark red region in FIG. \ref{fig:phase_diagram_isothermal_base}, i.e. the wind is subsonic when it enters the polytropic layer and is accelerated to  supersonic speeds in the polytropic layer. This case corresponds to the green curve in  panel (a) of FIG. \ref{fig:U_R_base}.



\begin{figure}
    \centering
    \includegraphics[width=\hsize]{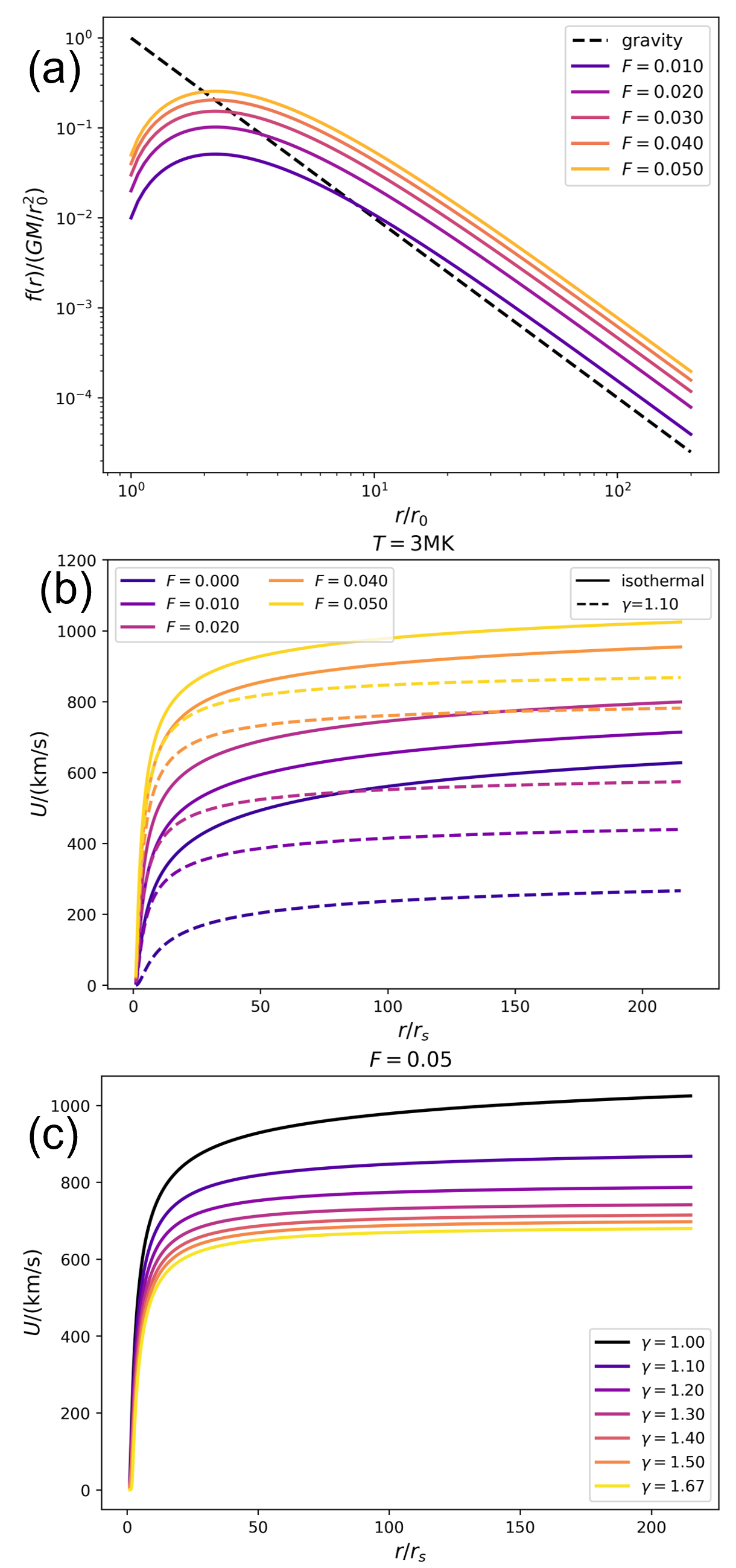}
    \caption{(a) Radial profile of the external force $f(r)$ (equation (\ref{eq:external_force})) with $\alpha=3$, $\beta=8$, and $f_0 = F  \times (GM/r_0^2)$. The black dashed line is the gravity force as a reference. (b) Radial profile of solar wind speed $V(r)$ with different external force strengths and $T_0=3$MK. Solid curves are isothermal case and dashed curves are $\gamma=1.1$. (c) Radial profile of solar wind speed $V(r)$ with $T_0=3$MK, $F=0.05$, and different $\gamma$.}
    \label{fig:external_force_U_R_isothermal}
\end{figure}

\subsection{External force}\label{sec:external_force}
In this section, we discuss the influence of the external force on the polytropic wind model. In the solar wind, Alfv\'en wave pressure gradient may provide such force as the amplitude of the wave magnetic field can be comparable to the mean magnetic field \cite{velli1991waves,kasper2019alfvenic,bale2019highly}. Rigorously, we need to model the amplitudes of outward and inward propagating Alfv\'en waves with two additional equations \cite{verdini2009turbulence,lionello2014validating,reville2020role,chandran2021approximate}. But these new equations will greatly complicate the system as they introduce a new critical point, the Alfv\'en point\cite{heinemann1980non,velli1993propagation}. Thus, in this section we assume the force $f(r)$ is a known function of $r$, so that the problem can be solved semi-analytically like we did in previous sections. The expression of $f(r)$ is
\begin{equation}\label{eq:external_force}
    f(r) = f_0  \frac{1 + \beta (s-1)}{s^3} \exp \left[\alpha\left(1- \frac{1}{s} \right)\right]
\end{equation}
where $\alpha$ and $\beta$ are two constants, $s = r/r_0$ and $f(r_0) = f_0$. Asymptotically, there is 
\begin{equation}
    f(r\rightarrow +\infty ) \rightarrow f_0\times  \frac{\beta  e^\alpha }{s^2},
\end{equation}
i.e., the external force decays as $r^{-2}$, consistent with a Wentzel-Kramers-Brillouin (WKB) decay of the Alfv\'en wave \cite{alazraki1971solar,belcher1971alfvenic,hollweg1974transverse}, which predicts an asymptotic $r^{-3/2}$ decay of the magnetic field fluctuations. 
The integral of $f(r)$ can also be written analytically:
\begin{equation}
\begin{aligned}
     \int f(r^\prime) dr^\prime = f_0 r_0  & \frac{\alpha \beta (s-1) + (1-\beta) s + \alpha}{\alpha^2 s} \\
     & \times \exp \left[\alpha\left(1- \frac{1}{s} \right)\right] + Const
\end{aligned}
\end{equation} 
We use one dimensionless parameter $F$ to measure the strength of $f_0$ such that $f_0 = F \times (GM/r_0^2)$. In this study, we fix $\alpha=3$, $\beta=8$, such that $f(r)$ peaks at $r=2.22r_0$, and we set $r_0 = r_s$. The radial profile $f(r)$ is plotted in panel (a) of FIG. \ref{fig:external_force_U_R_isothermal}, where the curves with different colors correspond to different values of $F$ and the black dashed line is the gravity force.
We note that the choice of the analytic form of $f(r)$, including values of $\alpha$ and $\beta$, is not fully rigorous. The only physics-based requirements are that the wave pressure gradient increases in the lower corona and asymptotically decays as $r^{-2}$. One can adjust the profile of $f(r)$, which will result in different profiles of the solution $V(r)$. But a parametric study of the influence of the shape of $f(r)$ on the solution $V(r)$ is beyond the scope of this study and is unnecessary for drawing the main conclusion of this section.

We start from the isothermal case ($\gamma = 1$). With the external force, the critical point at which $V=C = \sqrt{k_B T/ m_p}$, is determined by
\begin{equation}\label{eq:rc_isothermal_force}
    \left( \frac{GM}{r_c^2} - f(r_c) \right) \left( \frac{1}{A} \frac{dA}{dr} \right)^{-1}_{r_c} = C^2
\end{equation}
Obviously, adding an external force below the original sound point, i.e., the sound point without external force, does not change the location of the sound point nor the profiles of $V(r)$ beyond the sound point.
After numerically solving the critical point from equation (\ref{eq:rc_isothermal_force}), we can integrate the momentum equation from the critical point, which gives
\begin{equation}
\begin{aligned}
    \frac{1}{2} \left( V^2 - C^2\right) = C^2 \ln \left( \frac{VA}{C A_c} \right) & + GM \left( \frac{1}{r} - \frac{1}{r_c} \right) \\
    &+ \int_{r_c}^r f(r^\prime) d r^\prime
\end{aligned}
\end{equation}
to recover the profile of $V(r)$. We assume a spherical expansion such that $A(r) = r^2$. In panel (b) of FIG. \ref{fig:external_force_U_R_isothermal}, solid curves show the solar wind speed $V(r)$ with $T=3$MK and different $F$. It is clear that a stronger external force lowers the critical point and increases the solar wind speed significantly.

In the polytropic case, the closed equation set with non-zero $f(r)$ is similar to equation (\ref{eq:three_equation_set}):
\begin{subequations}\label{eq:three_equation_set_force}
\begin{equation}\label{eq:three_equation_force_critical_speed_and_sound_speed}
    V_c^2 = C_s^2 = C_0^2 \left(\frac{A_0 V_0}{A_c V_c} \right)^{\gamma-1}
\end{equation}
\begin{equation}\label{eq:three_equation_force_critical_speed_and_gravity}
    V_c^2 = \left(\frac{GM}{r_c^2} - f(r_c) \right) \times \left(\frac{1}{A} \frac{dA}{dr}\right)_{r_c}^{-1}
\end{equation}
\begin{equation}\label{eq:three_equation_force_integrated_mom_eq}
\begin{aligned}
     \frac{1}{2} (V_c^2 - V_0^2) = & - \frac{C_0^2}{\gamma - 1} \left[ \left( \frac{V_0A_0}{V_c A_c} \right)^{\gamma-1}  - 1\right] \\
     &+ GM \left(\frac{1}{r_c} - \frac{1}{r_0}\right) + \int_{r_0}^{r_c} f(r^\prime) dr^\prime
\end{aligned}
\end{equation}
\end{subequations}
From equation (\ref{eq:three_equation_force_critical_speed_and_gravity}), we get
\begin{equation}
    V_c^2 = \frac{C_g^2}{s_c^\prime}
\end{equation}
where we have defined a ``modified critical radius'' $s_c^\prime$ for convenience such that
\begin{equation}
    \frac{1}{s_c^\prime} = \frac{1}{s_c} - \frac{r_c f(r_c)}{2C_g^2} 
\end{equation}
Plug it into equation (\ref{eq:three_equation_force_critical_speed_and_sound_speed}), we get
\begin{equation}
    V_0^2 = C_g^2 \left( \frac{C_g}{C_0} \right)^{\frac{4}{\gamma-1}} \times s_c^4 \left( \frac{1}{s_c^\prime} \right)^{\frac{\gamma+1}{\gamma-1}}
\end{equation}
Then, plug $V_c$ and $V_0$ into equation (\ref{eq:three_equation_force_integrated_mom_eq}), we acquire the equation for $s_c$:
\begin{equation}\label{eq:sc_with_force}
    \begin{aligned}
         & \frac{1}{2} \left[ \frac{1}{s_c^\prime} - \left( \frac{C_g}{C_0} \right)^{\frac{4}{\gamma-1}} s_c^4 \left( \frac{1}{s_c^\prime} \right)^{\frac{\gamma+1}{\gamma -1}} \right] + \frac{1}{\gamma - 1} \left( \frac{1}{s_c^\prime} - \frac{C_0^2}{C_g^2} \right)\\
         &  = 2 \left( \frac{1}{s_c} - 1 \right) +  \frac{1}{C_g^2} \int_{r_0}^{r_c} f(r^\prime) dr^\prime
    \end{aligned}
\end{equation}
which is solved numerically. Finally, we integrate the montum equation from the critical point and get the profile $V(r)$. In panel (b) of FIG. \ref{fig:external_force_U_R_isothermal}, the dashed curves show $V(r)$ with $T_0=3$MK, $\gamma=1.10$ and varying $F$. Similar to the isothermal case, the external force increases the wind speed significantly. In addition, the external force weakens the constraint on the polytropic index $\gamma$. We have verified that a higher $F$ results in a larger range of $\gamma$ in which a solar wind solution can be found. For instance, with $T_0=3$MK, the solar wind solution is only possible for $\gamma \lesssim 1.15$ (FIG. \ref{fig:Nroots_sc_general}) without the external force. However, with $F \gtrsim 0.04$, $\gamma$ can be larger than $3/2$, and with $F \gtrsim 0.05$, $\gamma$ can take any value between $1$ and $5/3$. In panel (c), of FIG. \ref{fig:external_force_U_R_isothermal}, we plot $V(r)$ with $T_0=3$MK, $F=0.05$ and $\gamma$ varying from $1.0$ to $5/3$. Even with $\gamma=5/3$, the solar wind is still accelerated to over 600 km/s. 
However, we emphasize that the external force used here is not a self-consistent physics model. The selected values of $F$ lack comparison with observations as the radial profile of the wave pressure in the corona cannot be easily obtained from remote-sensing data. Assuming the wave amplitude at solar surface is $\delta u = 30$km/s, a typically value used in Alfv\'en wave models\cite{reville2020role}, and the scale of variation of the wave pressure is roughly $L=0.1r_s$, one can estimate $F\approx  (\delta u)^2/L / (GM_s/r_s^2) = 0.047$. Hence, a choice of $F=0.05$ may be reasonable, but future studies are necessary to quantify the Alfv\'en wave pressure in the solar corona and solar wind.


\section{Conclusion}\label{sec:conclusion}
In this study, we make use of PSP data from the first nine encounters to estimate the polytropic index of the solar wind proton. The radial profiles of the proton temperature indicate that the polytropic index is highly dependent on the solar wind speed. Faster solar wind in general has smaller polytropic index than the slower solar wind. Solar wind stream faster than 400 km/s may have a polytropic index around 1.25 while the stream slower than 300 km/s may have a polytropic index around $5/3$ (FIG. \ref{fig:PSP_data}).

We then carry out a comprehensive analytic-numerical study of the 1D polytropic solar wind model. The major results of this part are summarized below:
\begin{enumerate}
    \item For a generic star with $C_g$($=\sqrt{GM/2r_0}$), $C_0$(=$\sqrt{\gamma k_B T_0/m_p}$), and polytropic index $\gamma$, the accelerating transonic stellar wind solution only exists in the parameter space bounded by $C_0/C_g < 1$ and $(C_0/C_g)^2 >2(\gamma-1)$. Naturally, there is a constraint in $\gamma$ such that $1\leq \gamma <3/2$.
    \item For the Sun, with realistic surface temperature $T_0=1 \sim 3$ MK, no solar wind solution exists with the polytropic index $\gamma \gtrsim 1.25$ which is deduced from in-situ measurements.
    \item An isothermal layer, in which the temperature remains constant, may help generate a solar wind with large $\gamma$. But the upper boundary of the layer must not be too close to the isothermal critical radius in order to produce a continuously accelerated solar wind. Whether there is such an isothermal layer is a question and needs further studies using in-situ measurements as close as possible to the Sun and remote-sensing observations.
    \item The external force, e.g. the Alfv\'en wave pressure gradient, may also contribute to overcome the constraint in $\gamma$.
\end{enumerate}
These results indicate that both the lower coronal heating by thermal conduction or processes like magnetic reconnection, and the in-situ dynamics such as the Alfv\'en waves can be very important in generation of the observed polytropic solar wind. 

\begin{acknowledgments}
This work is supported by NASA HERMES DRIVE Science Center grant No. 80NSSC20K0604 and the NASA Parker Solar Probe Observatory Scientist grant NNX15AF34G.
\end{acknowledgments}

\section*{Data Availability Statement}
The Parker Solar Probe data used in this study are available to public on NASA's Coordinated Data Analysis Web (CDAWeb) (\href{https://cdaweb.gsfc.nasa.gov/pub/data/psp/}{https://cdaweb.gsfc.nasa.gov/pub/data/psp/}). The numerical results presented in the paper are available upon reasonable request from the corresponding author.

\appendix*
\section{Existence of transonic solutions}\label{sec:append_Nroot}
\subsection{Number of roots}
In this appendix, we analyze equation (\ref{eq:sc_no_force}) in details and explain the method to determine whether the equation has roots and how many roots exist with given parameters. For convenience, let us define $x = 1/s_c$, $a = (C_0/C_g)^2$, and $b(\gamma) = (5-3\gamma)/(\gamma - 1)$ so that equation (\ref{eq:sc_no_force}) can be reformed into
\begin{equation}\label{eq:app_sc_reform}
    E(x) = a^{-\frac{2}{\gamma - 1}} \cdot x^b - b\cdot x + \left( \frac{2a}{\gamma - 1} - 4 \right) = 0.
\end{equation}
We aim to find the solutions in the range $x\in(0,1]$ with free parameters $a \in (0,+\infty)$ and $\gamma \in (1,5/3]$, which means $b \in \left[0, +\infty \right)$. Especially, $b(3/2) = 1$ and $b(5/3) = 0$.

First, we can write down 
\begin{equation}
    E(0) = \frac{2a}{\gamma-1} - 4, \quad E(1) = a^{-\frac{2}{\gamma - 1}}  +  \frac{2a-\gamma-1}{\gamma - 1} 
\end{equation}
By taking the derivative of $E(x)$:
\begin{equation}
    E^\prime(x) = b \left( a^{- \frac{2}{\gamma - 1}} x^{b-1} - 1\right),
\end{equation}
we see that there is \textit{only one extremum} ($E^\prime(x_e) = 0$) within $x \in (0,+\infty)$, at $x_e = a^{\frac{1}{3-2\gamma}}$, and 
\begin{equation}
    E(x_e) =  \frac{2(2\gamma -3)}{\gamma - 1} a^{\frac{1}{3-2\gamma}} + \frac{2a}{\gamma -1 } - 4
\end{equation}
Thus, given the values of $a$ and $\gamma$, the number of roots can be determined with the following process:
\begin{enumerate}
    \item Calculate $E(0)$, $E(1)$, $x_e$, and $E(x_e)$
    \item If $E(0)E(1) < 0$, there is only \textit{one root}. If $E(0)E(1) > 0$, go to the next step.
    \item If $x_e \geq 1$, there is \textit{no root}. If $0<x_e<1$, go to the next step.
    \item If $E(0)E(x_e) > 0 $, there is \textit{no root}. If $E(0)E(x_e) < 0 $, there are \textit{two roots}. If $E(x_e) = 0$, there is \textit{one root}.
\end{enumerate}
To complete the analysis, we need to discuss the case $E(0) E(1) = 0 $. 

If $E(0) = 0$, we get $a = 2(\gamma - 1 )$ 
and thus
\begin{equation}
    E(1) = \left[ 2(\gamma - 1) \right]^{-2/(\gamma-1)} - b.
\end{equation}
One can show that $E(1) \geq 0$ for $\gamma \in (1,5/3]$ with the zero point at $\gamma = 3/2$, which is a special condition and will be discussed later. Here let us ignore the $\gamma=3/2$ case and write $E(1)>0$. 
\begin{equation}
    x_e = \left[2(\gamma - 1) \right]^{1/(3-2\gamma)},
\end{equation}
which is always smaller than $1$ with $\gamma \in (1,5/3]$. $E(x_e)$ is negative for $1<\gamma<3/2$ and positive for $\gamma > 3/2$. Thus, along the line $a=2(\gamma -1)$, there is one root for $1 < \gamma < 3/2$ and no root for $3/2 < \gamma < 5/3$. 

If $E(1) = 0$, one can show that $a=1$ is the only possibility for any $\gamma$ value. Then we have $x_e= 1$. Thus, no matter whether $E(0)$ is negative or positive (separated by $\gamma=3/2$), there is only one root along the $a=1$ line, which is $x=1$ such that the flow starts with sound speed at the inner boundary.

One special case is $\gamma = 3/2$ ($b=1$) when $E(x)$ is a linear function of $x$. In this case, one can show that
\begin{equation}
E(0) E(1)= 4(a-1)^2 \left[4- (a+1)(a^2+1)/a^4 \right].
\end{equation}
For $a<1$, $E(0)E(1) < 0$ and there is one root. For $a>1$, $E(0)E(1) > 0$ and there is no root. For $a=1$, we get $E(x) \equiv 0$, i.e. any $x$ is allowed. By observing the integrated momentum equation (equation (\ref{eq:mom_integral})), we find that with $\gamma=3/2$ and $a=1$, $V(r) \equiv V_0$ is an exact solution. That is to say, the flow can start from any initial value and remains a constant speed. That is why any critical point is allowed. 

Another case is $\gamma = 5/3$ ($b=0$) and all terms containing $x$ vanish in equation (\ref{eq:app_sc_reform}), leading to the equation
\begin{equation}
    a^{-3} + 3a = 4
\end{equation}
The equation is not satisfied in general unless $a=1$, when $x$ can be of any value. Actually, one can show that with $\gamma=5/3$ and $a=1$, $V(r) = C_0 s^{-1/2}$ is a solution, and the wind speed equals to the sound speed ($V(r) = C_s(r)$) at any location $r$. Hence, for adiabatic plasma, there is no transonic solution at all. 

The phase diagram of the number of transonic solutions on the $a-\gamma$ plane is shown in FIG. \ref{fig:Nroots_sc_general}.

\subsection{Property of the solutions}
We note that, so far we only determined the number of roots. For each root, whether the solution $V(r)$ is accelerating or decelerating with $r$ needs further analysis. 
The simplest way is to calculate $V_0$ (equation (\ref{eq:relation_V0_and_rc})) and $V_c$ (equation (\ref{eq:relation_Vc_and_rc})) after solving $r_c$ and compare the two values. If $V_c > V_0$ the flow is accelerating and vice versa. 

We do not find a straightforward way to make such comparison analytically (see next paragraph for a weak proof with both $a<2(\gamma-1)$ and $a<1$ satisfied), thus we numerically calculate these values. We have verified that if there is only one root of $r_c$, only the decelerating solution ($V_c<V_0$) exists, and if there are two roots of $r_c$, one of the root corresponds to $V_c<V_0$ and the other one corresponds to $V_c>V_0$. That is to say, the accelerating solution ($V_c>V_0$) only exists in the dark red region shown in FIG. \ref{fig:Nroots_sc_general} that is bounded by $a<1$ and $a>2(\gamma - 1)$.

In this paragraph, we provide a proof that if $a<2(\gamma-1)$ and $a<1$ are both satisfied, i.e. the orange region in FIG. \ref{fig:Nroots_sc_general} except for the triangle above $a=1$, the solution must have $V_c<V_0$. We can write equations (\ref{eq:relation_V0_and_rc}) and (\ref{eq:relation_Vc_and_rc}) as 
\begin{equation}
    \frac{V_0^2}{C_0^2} = a^{-(\gamma+1)/(\gamma - 1)} x_c^b 
\end{equation}
and 
\begin{equation}
    \frac{V_c^2}{C_0^2} = \frac{x_c}{a}
\end{equation}
where $x_c = 1/s_c$. Thus,
\begin{equation}
    \frac{V_c^2}{V_0^2} = \frac{x_c}{a^{-2/(\gamma - 1)} x_c^b} = \frac{x_c}{b x_c + 4 - \frac{2a}{\gamma - 1}}
\end{equation}
where equation (\ref{eq:app_sc_reform}) is used to eliminate $x^b$. Consider $a<2(\gamma  - 1)$, i.e. the orange region in FIG. \ref{fig:Nroots_sc_general}, such that the denominator is always positive (note that $b>0$ and $x_c>0$). If we assume $V_c^2/V_0^2 > 1$, we get
\begin{equation}
    (1-b)x_c > 4 - \frac{2a}{\gamma - 1}
\end{equation}
For $1< \gamma < 3/2$, $1-b < 0$, but $x_c>0$ and $4- 2a/(\gamma - 1) > 0$, so the above inequality cannot be satisfied. Thus, in FIG. \ref{fig:Nroots_sc_general}, the orange region on the left of $\gamma = 3/2$ can only have a decelerating solution. For $\gamma > 3/2$, $1-b>0$ and thus we get 
\begin{equation}
    x_c > 1 + \frac{1-a}{2 \gamma - 3}
\end{equation}
Obviously, for $a<1$, the above inequality means $x_c > 1$, which is out of the domain ($0<x \leq 1$). Thus, we have proven that the orange region on the right of $\gamma =3/2$ and below $a=1$ can only have a decelerating solution. We note that no analytic proof for the triangle bounded by $a<2(\gamma-1)$ and $a>1$ is found but we have numerically verified that in this region $V_c<V_0$ is still valid.

An interesting point is that, below $a=1$ line, the decelerating solution starts at $r_0$ with supersonic speed, i.e. $V_0/C_0>1$ and crosses the sound point from above. However, above $a=1$ line (and below $a=2(\gamma-1)$), the decelerating solution starts with subsonic speed, i.e. $V_0/C_0<1$, and crosses  the sound point from below. That is to say, both $C_s(r)$ and $V(r)$ are decreasing but $C_s(r)$ decreases faster than $V(r)$ so that they cross at the sound point. It means that, although $V(r)$ is a decreasing function, the flow transits from subsonic to supersonic as it propagates away from the star.

\bibliography{references}

\end{document}